\documentclass [pra, aps, groupedaddress,twocolumn,floatfix] {revtex4}
\usepackage{graphicx}
\usepackage{amsmath}
\usepackage{amssymb}
\usepackage{color}

\begin{document}
\title{Collective modes of a two-dimensional spin-$1/2$ Fermi gas in a harmonic trap}
\author{Stefan K. Baur}
\affiliation{T.C.M. Group, Cavendish Laboratory, J. J. Thomson Avenue, Cambridge CB3 0HE, United Kingdom}
\author{Enrico Vogt}
\affiliation{AMOP Group, Cavendish Laboratory, J. J. Thomson Avenue, Cambridge CB3 0HE, United Kingdom}
\author{Michael K\"ohl}
\affiliation{AMOP Group, Cavendish Laboratory, J. J. Thomson Avenue, Cambridge CB3 0HE, United Kingdom}
\author{Georg M. Bruun}
\affiliation{Department of Physics and Astronomy, University of Aarhus, Ny Munkegade, DK-8000 Aarhus C, Denmark}

\date{\today}

\begin{abstract}
We derive  analytical expressions for the frequency and damping of the lowest collective modes of a two-dimensional 
Fermi gas using kinetic theory. For strong coupling, 
we furthermore show that pairing correlations overcompensate the effects of Pauli blocking on the collision rate for a large range of temperatures, resulting 
in a rate which is larger than that of a classical gas. 
Our results agree well with experimental data, and they recover the observed cross-over from collisionless to hydrodynamic 
behavior with increasing coupling for the quadruple mode. Finally, we show that a trap anisotropy within the experimental bounds results in
 a  damping of the breathing mode which is comparable to what is observed,  even for a scale invariant system. 
\end{abstract}
\maketitle
\section{Introduction}

A new generation of experiments realizing  two-dimensional (2D) atomic Fermi gases provide a unique possibility 
to systematically explore 2D quantum systems, including phenomena relevant to high temperature and organic superconductors, and $^3$He 
films~\cite{Martiyanov:2010,Frohlich:2011cr, Dyke:2011dz,Sommer:2012fv}.  
The study of collective modes has proven to be a  powerful 
way to probe the properties of 3D atomic gases including the effects of pairing and strong correlations~\cite{Chevy:2002ly,Riedl:2008bh,Giorgino:2008,Khoon-Tey:2012kx}.
The damping of these modes is related to the transport properties of the gases, and 
in particular the damping of the quadrupole mode is controlled by the shear viscosity in the hydrodynamic limit.
 The shear viscosity of quantum Fermi fluids has received lots of 
interest due to the conjectured minimum of the ratio of viscosity and entropy~\cite{Kovtun:2005qa,Schafer:2007mi,Cao:2011kl}. 
Recently, the first experiments  concerning the collective modes of a 2D Fermi gas have been performed~\cite{Vogt:2012vn}. 
The breathing mode frequency was reported to be very close to that of an ideal gas with a low and constant damping throughout the different interaction regimes, which indicates a 
scale invariant system with a very small bulk viscosity~\cite{Olshanii:2010zr,Hofmann:2012fk,Taylor:2012kx,Gao:2012uq}. The quadrupole mode was shown to exhibit a clear transition between collisionless and 
hydrodynamic behavior with increasing coupling strength. Following this experiment, the shear viscosity for a 2D  gas was calculated using kinetic 
theory, and the result was then used to calculate the quadrupole mode damping in the hydrodynamic limit~\cite{Bruun:2012dq,Schafer:2012cr,Enss:2012ly}. 
From the good agreement between experimental data for the quadrupole mode and a numerical solution of the Boltzmann equation,
it was concluded that kinetic theory provides an accurate description in the temperature and coupling strength regime relevant to the experiment~\cite{Wu:2012zr}. 
Here, we calculate the frequency and damping of the energetically lowest collective  modes of a 2D Fermi gas as a function of temperature  
and coupling strength from  approximate
 analytic solutions to the  Boltzmann equation obtained by expanding it in appropriate basis functions. The effects of Pauli blocking and pairing correlations on the collision rate are systematically analyzed, and we show that for strong coupling  there is a large temperature regime where pairing correlations dominate the effects of Pauli blocking. This indicates a large so-called pairing pseudo-gap region in the Bose-Einstein condensate (BEC) to Bardeen-Cooper-Schrieffer (BCS) phase diagram~\cite{Randeria:1992bh,Trivedi:1995dq,Perali:2002cr,Stajic:2004fk,Chen:2005uq,Feld:2011nx}. Our predictions for the mode frequencies and damping are 
shown to agree well with the experimental results. In particular, the theory reproduces quite well the collisionless to hydrodynamic cross-over of the quadrupole mode with increasing coupling strength. By including the effects of a possible slight trap anisotropy, we show that the resulting coupling between quadrupole and breathing modes 
can lead to additional damping of the breathing mode even when the bulk viscosity is zero. This result can be important for a proper interpretation of experiments. \\
Our paper is  organized as follows: First we derive the collective mode frequencies using the method of moments. Then we proceed by calculating
the collision rate using various levels of approximations. We demonstrate that the collision
rate can be larger than that of a classical gas due to pairing correlations. Our results are then shown to provide a good 
quantitative description of the experimentally observed collective modes. Finally, we explore how a  trap anisotropy within the experimental bounds results in a 
significant damping of the breathing mode. 
\begin{figure*}
\includegraphics[width=2.05\columnwidth]{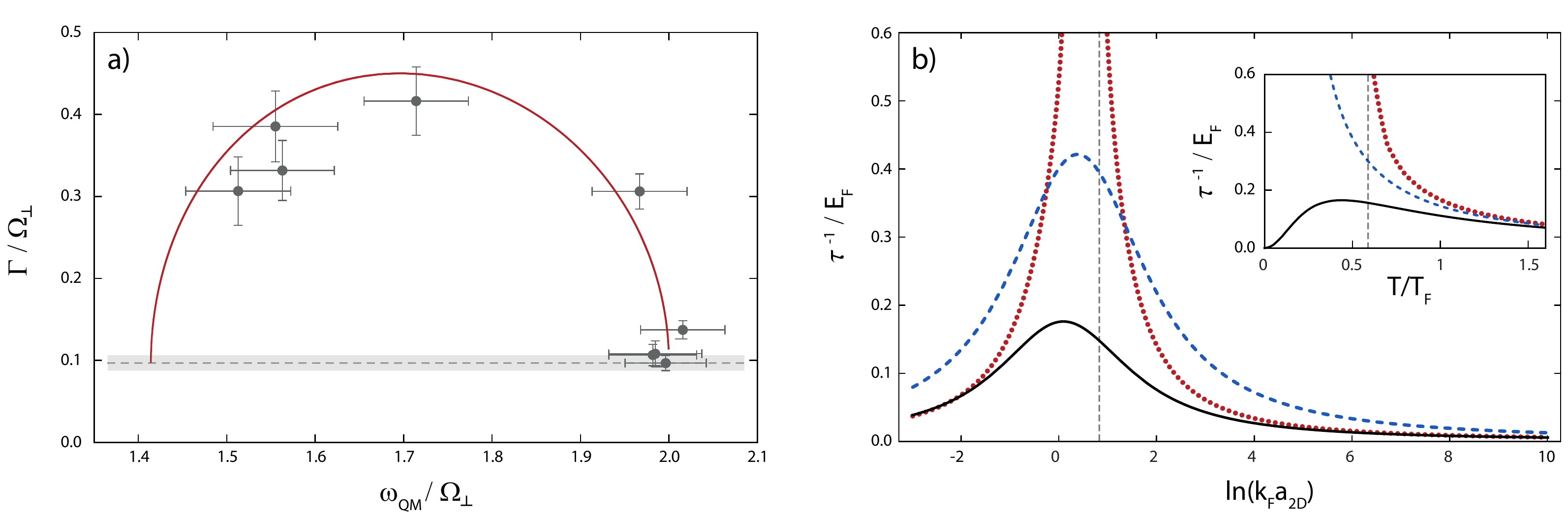}
\caption{(Color online) (a) Plot of damping $\Gamma$ versus frequency $\omega_{\rm QM}$ for the quadrupole mode. The theory curve (solid line) relies only on the moment method and is independent of a microscopic model for the collision time $\tau$. The dots are the experimental data. (b) The collision rate $1/\tau$ as a function of interaction strength at $T/T_F=0.47$. The inset shows the relaxation rate as a function of $T$ at fixed interaction strength $\ln (k_F a_{\rm 2D})=0.5$. 
All theory curves were calculated with (solid black line) and without (dashed blue line) 	 and with Pauli blocking and medium corrections (red dotted line).
 The collision rate including medium corrections diverges when 
   $T^*$ as given by the Thouless criterion (indicated by the vertical dashed line) is approached from above. 
}
\label{fig:fig1}
\end{figure*}
\section{Formalism}
We consider a gas of $N$ fermionic atoms of mass $m$ trapped in a  harmonic potential
\begin{eqnarray}
V(x,y,z)=\frac{1}{2} m \left( \omega_x^2 x^2+ \omega_y^2 y^2+ \omega_z^2 z^2 \right).
\end{eqnarray}
 There is an equal number $N/2$ of atoms in two internal 
spin states which we denote $\uparrow$ and $\downarrow$, and atoms in different spin states  interact via a short-range potential characterized by a 
scattering length $a$ whereas the interaction between equal spin atoms can be neglected.  We focus on the 
quasi-2D limit with $\omega_x, \omega_y \ll \omega_z$ and $ T, E_F \ll  \omega_z$, where $T$ is the temperature 
and   $T_F\equiv E_F=\sqrt{N \omega_x \omega_y}$ the Fermi temperature  (for brevity, we  set $\hbar=k_B=1$).
 In this limit, the motion of the atoms is frozen along the $z$-axis and the kinematics of the system is essentially 2D.  

We use the semi-classical Boltzmann equation 
\begin{eqnarray}
\label{eq:boltzmann}
\left(\partial_t+\dot{\mathbf{r}}\cdot \nabla_{\mathbf{r}}+\dot{\mathbf{p}} \cdot \nabla_{\mathbf{p}} \right)f=-I[f]
\end{eqnarray}
to describe the collective motion of the gas, where $f(\mathbf{r},\mathbf{p})$ is the distribution function,
 $\dot{\mathbf{r}}=\mathbf{p}/m$ and $\dot{\mathbf{p}}=-
\nabla_{\mathbf{r}} V$. Since we consider density modes, we have taken the same distribution function for the   
$\uparrow$ and the $\downarrow$ atoms.
As in Refs.~\cite{Bruun:2012dq,Schafer:2012cr,Enss:2012ly}, we assume that the main effects of interactions are
 captured by the collision integral $I$ so that interactions effects on the left side of  the Boltzmann equation can be 
 ignored~\cite{Enss:2012ly}.
The collective modes can  be obtained by linearizing  the Boltzmann equation around the equilibrium distribution function
\begin{eqnarray}
\label{eq:equilibrium}
f_0(\mathbf{r},\mathbf{p})=\frac{1}{e^{\beta[\epsilon_{\mathbf{p}}+V(\mathbf{r})-\mu]}+1},
\end{eqnarray}
where $\beta=1/T$ and $\epsilon_{\mathbf{p}}=\mathbf{p}^2/2m$. In order to linearize Eq.~\eqref{eq:boltzmann}, we write
 $f(\mathbf{r},\mathbf{p})=f_0(\mathbf{r},\mathbf{p})+f_0(\mathbf{r},\mathbf{p}) \left[1-f_0(\mathbf{r},\mathbf{p})\right] \phi(\mathbf{r},\mathbf{p})$, 
 where the factor $f_0 \left(1-f_0\right)$ is introduced for later convenience~\cite{PethickSmith}.
  The linearized Boltzmann equation then takes the form
\begin{eqnarray}
\label{eq:linearizedboltzmann}
f_0 (1-f_0) \left(\partial_t+\dot{\mathbf{r}}\cdot \nabla_{\mathbf{r}}+\dot{\mathbf{p}} \cdot \nabla_{\mathbf{p}} \right) \phi=-I[\phi]
\end{eqnarray}
with the collision integral  
\begin{gather}
I[\phi_1]=\int d^2 \check p_2\frac{m_r}{2\pi}\int_0^{2\pi}\!\! d\theta'|{\mathcal T}|^2(\phi_1+\phi_2-\phi_3-\phi_4)
\nonumber\\
f_1f_2(1-f_3)(1-f_4)
\label{CollisionIntegral}
\end{gather}
where ${\phi}_1$ stands for $\phi(\mathbf{r},{\mathbf{p}}_1)$ etc., and we have defined $ d^2 \check p= d^2 p/(2\pi)^2$. The scattering  
of particles with momenta $\mathbf{p}_1$ and $\mathbf{p}_2$ to momenta $\mathbf{p}_3$ and $\mathbf{p}_4$ with ${\mathbf p}_4={\mathbf p}_1+{\mathbf p}_2-{\mathbf p}_3$
is described by the scattering matrix ${\mathcal T}$. The relative mass is $m_r=m/2$ and $\theta'$ is 
the angle between the outgoing relative momentum ${\mathbf p}_r'=({\mathbf p}_3-{\mathbf p_4})/2$ and the  
center of mass momentum ${\mathbf P}={\mathbf p}_1+{\mathbf p_2}$. Note that ${\mathbf P}$ is conserved in the collision and that 
 $|{\mathbf p}_r|=|{\mathbf p}_r'|$ with  ${\mathbf p}_r=({\mathbf p}_1-{\mathbf p_2})/2$.
 In Eq.~(\ref{CollisionIntegral}) and the following, all distribution functions  are  of the equilibrium form 
 given by Eq.~(\ref{eq:equilibrium}).

In vacuum, a collision between a $\uparrow$ and a $\downarrow$ atom with momentum ${\mathbf p}_r$ and kinetic 
energy $\epsilon=p_r^2/2 m_r$ in the center of mass frame is described by the low energy 2D ${\mathcal T}$-matrix~\cite{landaulifshitz}
\begin{eqnarray}
\label{eq:vact}
{\cal T}_{\rm v}(\epsilon)=\frac{2\pi}{m_r} \frac{1}{\ln(\epsilon^*/\epsilon)+i \pi}
\end{eqnarray}
where  $\epsilon^*=\hbar^2/m a_{\rm 2D}^2$ with the 2D scattering length $a_{\rm 2D}=l_z \sqrt{\pi/B}\; e^{-\sqrt{\pi/2} l_z/a_s}$, with $l_z=1/\sqrt{m\omega_z}$ and $B=0.905$~\cite{Petrov:2001fk}.

\begin{figure*}
\includegraphics[width=2.05\columnwidth]{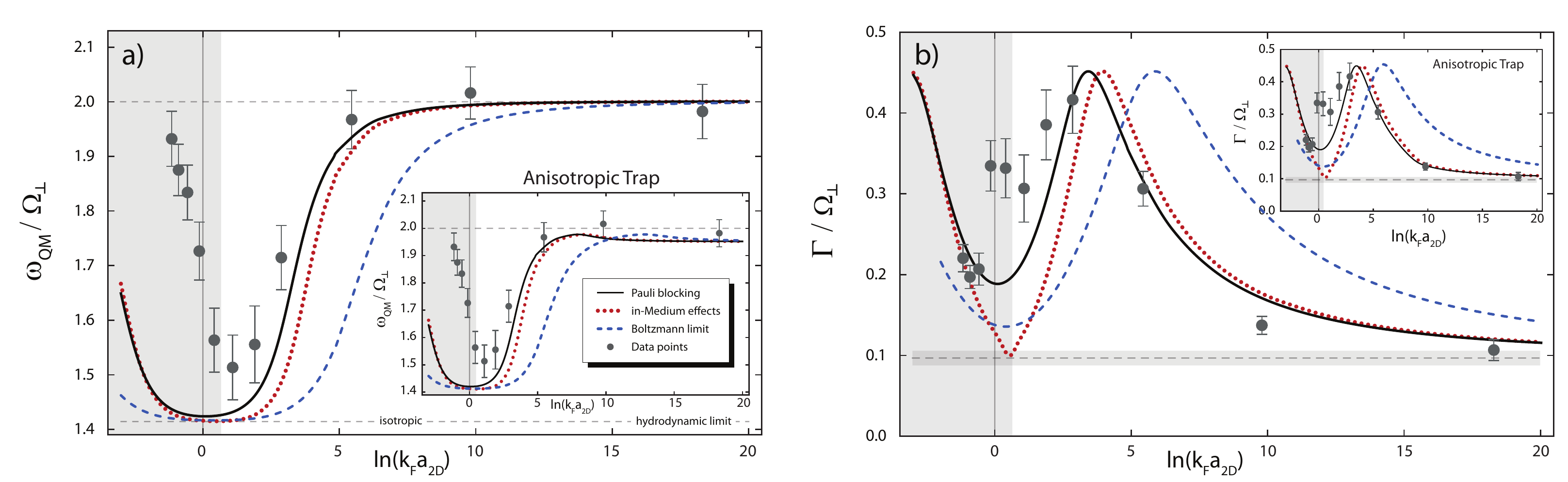}
\caption{(Color online) Quadrupole mode frequency (a) and damping (b) in units of $\Omega_{\bot}=(\omega_x\omega_y)^{1/2}$ compared to the experimental data of Ref.~\cite{Vogt:2012vn}.  In the gray shaded area, the normal state is unstable towards pairing (within BCS mean-field theory). We have added an offset of $0.1\Omega_{\perp}$ to all damping theory curves to account for other sources of damping present in the experiment of Ref.~\cite{Vogt:2012vn}. The insets show the eigenmode frequencies and damping rates for an anisotropic trap ($\omega_y=1.05\,\omega_x$), which leads to a better agreement between theory and experimental data.  
}
\label{fig:fig2}
\end{figure*}
The collective modes are solutions to Eq.\ (\ref{eq:linearizedboltzmann}) of the form 
 $\phi(\mathbf{r},\mathbf{p},t)=e^{-i\omega t} \phi(\mathbf{r},\mathbf{p})$ where ${\rm Re}(\omega)$ is the mode frequency and 
 ${\rm Im}(\omega)$ is the damping rate.  We 
 expand $\phi(\mathbf{r},\mathbf{p})$ on a set of basis functions which form a closed set 
 under application of the left hand side of Eq.~\eqref{eq:linearizedboltzmann}, i.e.\ we write 
 $\phi(\mathbf{r},\mathbf{p})=\sum_lc_l \phi_l(\mathbf{r},\mathbf{p})$.
  Several previous studies have applied this method (or variations of it) successfully to harmonically trapped gases (see e.g.~\cite{PethickSmith, Guery-Odelin:1999kx,  Ghosh:2000vn,Bruun:2007qf,Chiacchiera:2011zr}),
and here  we apply it to a 2D  Fermi gas.

\section{Isotropic trap}
The trap used in the experiment in Ref.~\cite{Vogt:2012vn} is nearly isotropic in the $xy$-plane, and we therefore first consider the case
 $\omega_x=\omega_y\equiv\Omega_{\perp}$. 
 
For the breathing mode,  we use the basis set 
 \begin{equation}
\label{eq:basis1}
\phi_1=\Omega_{\perp}^2 (x^2+y^2),\, \phi_2=\Omega_{\perp} (x p_x+y p_y),\, \phi_3=p_x^2+p_y^2.
\end{equation}
Since this set is closed under application of the left-hand side of Eq.~\eqref{eq:linearizedboltzmann} and the 
collision integral  vanishes by conservation of energy and momentum, one finds an undamped breathing mode at 
$\omega_B=2 \Omega_{\perp}$~\cite{Guery-Odelin:1999kx,Goulko:2012nx}. 
 This is a consequence of harmonic trapping and rotation symmetry,
  and the result agrees surprisingly well with what is seen in experiments~\cite{Vogt:2012vn}.

For the quadrupole mode, we use the basis set 
\begin{equation}
\label{eq:basis2}
\phi_1=\Omega_{\perp}^2 (x^2-y^2), \, \phi_2=\Omega_{\perp} (x p_x-y p_y), \, \phi_3=p_x^2-p_y^2.
\end{equation}
The corresponding equation determining the mode frequency is~\cite{Riedl:2008bh}
\begin{eqnarray}
 \label{eq:qpmode}
 \omega^2-\omega_{\rm cl}^2+\frac{i}{\omega \tau} (\omega^2-\omega_{\rm hd}^2)=0
\end{eqnarray}
where $\omega_{\rm cl}=2 \Omega_{\perp}$ is the collisionless and $\omega_{\rm hd}=\sqrt{2} \Omega_{\perp}$ the 
hydrodynamic limit of the quadrupole mode frequency. 
Indeed, Eq.~(\ref{eq:qpmode}) has the solution 
 \begin{equation}
\omega=\omega_{\rm cl}-\frac i{4\tau}
\end{equation}
in the collisionless limit $\omega\tau\gg 1$, and the solution 
 \begin{equation}
\omega=\omega_{\rm hd}-i\tau \Omega_{\perp}^2.
\label{hydrosolution}
\end{equation}
in the hydrodynamic limit $\omega\tau\ll 1$. This  explicitly demonstrates the different roles of collisions in the two regimes. 
 The effective collision rate $1/\tau$ which comes from the non-zero collision integral of $\phi_3$, is
\begin{eqnarray}
\label{eq:relaxationtime}
\frac{1}{\tau}=\frac{\int d^2r d^2 \check{p}\;(p_x^2-p_y^2) I[p_x^2-p_y^2]}{\int d^2r d^2\check{p}\; f_0(1-f_0) (p_x^2-p_y^2)^2}.
\end{eqnarray}

Before using Eq.~(\ref{eq:relaxationtime}) to calculate $\tau$ as a function of temperature and interaction strength, we  first eliminate $\tau$
from Eq.~(\ref{eq:qpmode}) to obtain $\Gamma_Q=\sqrt{\sqrt{8(\omega_Q \Omega_{\perp})^2-7 \Omega_{\perp}^4}-\Omega_{\perp}^2-\omega_Q^2}$
where $\Gamma_Q={\rm Im}(\omega)$ is the damping of the quadrupole mode with frequency $\omega_Q={\rm Re}(\omega)$~\cite{Riedl:2008bh}.
In Fig.\ref{fig:fig1} (a),
we compare this with the experimental data of Ref.~\cite{Vogt:2012vn} by plotting the damping rate as a function of the frequency.
The  good agreement indicates that our approach for calculating the collective modes captures the most important physics of the experiment. 
In particular, it shows that extending the basis set given in Eq.~(\ref{eq:basis2}) to include higher order functions~\cite{Chiacchiera:2011zr}
 will  most likely only  yield a fairly modest gain in accuracy.

\section{Collision rate}
 To make further connection to experiments, we must calculate $\tau$  as a function of  temperature and interaction strength. 
It is easiest to calculate $\tau$ in the high-temperature classical limit $z=\exp(\beta\mu)\ll 1$, where 
we can ignore the Fermi-blocking factor $(1-f_3)(1-f_4)$ in Eq.~(\ref{CollisionIntegral}) and use the  
vacuum ${\mathcal T}$-matrix given by Eq.~(\ref{eq:vact}). Equation \eqref{eq:relaxationtime} then becomes
\begin{equation}
\frac{1}{\tau_{\rm cl}}=\frac{\pi z T}{2}  G(\epsilon^*/T),
\label{tauclass}
\end{equation}
with the dimensionless integral
\begin{eqnarray}
G(y)=\int_0^{\infty}  \hspace{-2mm} dt\; \frac{t^2 e^{-t}}{\ln^2(y/t)+\pi^2}.
\end{eqnarray}
Converting the fugacity $z$ to particle number using the ideal gas formula $N=N_{\uparrow}+N_{\downarrow}=2 T^2z/\omega_x \omega_y$ gives 
\begin{equation}
\frac{1}{\tau_{\rm cl}}=\frac{\pi N \omega_{x} \omega_y}{4 T} G(\epsilon^*/T).
\end{equation}
Comparing this expression for the collision rate in a  harmonically trapped gas in the classical limit with that for the 
shear viscosity $\eta_{\rm cl}$ of a uniform gas in the classical limit~\cite{Bruun:2012dq, Schafer:2012cr}, we get
\begin{eqnarray}
 \eta_{\rm cl}=\frac T 2 n(0)\tau_{\rm cl}
 \label{etalink}
\end{eqnarray}
where $n(0)=mN\omega_x\omega_y/2\pi T$ is the density in the center of the trap. This is the usual relation between the collision time and the 
viscosity in the classical limit, with the factor $1/2$ reflecting the trap average. From Eqs.~(\ref{hydrosolution}) and (\ref{etalink}),
we furthermore  see that the 
damping is proportional to the shear viscosity in the hydrodynamic limit as expected~\cite{LandauLifshitzFluid,Bruun:2012dq, Schafer:2012cr}.

For lower temperatures, the Pauli blocking factor $(1-f_3)(1-f_4)$ in Eq.~(\ref{CollisionIntegral})
 for the collision integral significantly decreases the collision rate. On the other hand, medium effects  in the scattering matrix were found to essentially 
cancel this reduction due to the pairing instability for a 3D gas in the unitarity limit, resulting in a  collision rate  close to the classical 
prediction for temperatures all the way down to the critical temperature $T_c$ 
for superfluidity~\cite{Riedl:2008bh}. In 2D, medium effects on the scattering matrix were 
even found to over-compensate the effects of Pauli blocking when calculating the shear viscosity 
and spin diffusion coefficient in the strong coupling limit~\cite{Enss:2012ly}. In addition to keeping the Pauli blocking terms in 
Eq.~(\ref{CollisionIntegral}) for lower temperatures, we therefore also include medium 
effects in the ${\mathcal T}$-matrix. This gives in the ladder approximation~\cite{Schmitt-Rink:1989uq}
\begin{equation}
\label{eq:mbtmatrix}
\frac1{{\cal T}_{\rm MB}(P,\epsilon)}=\frac1{{\cal T}_{\rm v}(\epsilon)}+
\int d^2\check k \frac{f_{\mathbf{k}+\mathbf{P}/2}+f_{\mathbf{k}-\mathbf{P}/2}}{\epsilon+i0-k^2/m-P^2/4m},
\end{equation}
where $i0$ denotes an infinitesimal positive imaginary part. 
Because center-of-mass and relative coordinates are coupled in the presence of a Fermi sea, this adds an extra dependence of the ${\mathcal T}$-matrix
 on the center-of-mass momentum ${\mathbf P}$. Fortunately, the numerical evaluation $1/\tau$ can be simplified significantly by using the symmetry or anti-symmetry of the 
  collision operator under the exchange of its momenta, and the invariance under a  
  simultaneous rotation of all momenta. Using this, the numerator  of  
 Eq.~(\ref{eq:relaxationtime}) can be simplified to 
   \begin{gather}
\frac{m_r}{4\pi^2}\int d^2r\int_0^\infty\!\! dPP\int d^2 \check p_r p_r^4\int_0^{2\pi}\!\!d\theta'
|{\mathcal T}_{\rm MB}|^2\nonumber\\
[1-\cos(2\theta-2\theta')]f_1f_2(1-f_3)(1-f_4)
\label{TauSym}
\end{gather}
 where $\theta$ is the angle  between ${\mathbf p}_r$  and  ${\mathbf P}$. 
 The denominator in (\ref{eq:relaxationtime}) can be written as 
  \begin{equation}
\int d^2rd^2\check p  f_0(1-f_0)(p_{x}^2-p_{y}^2)^2=-\frac{4 T^4 m^2}{\omega^2} {\rm Li}_3(-z)
\end{equation}
where ${\rm Li}_s(x)$ is the polylogarithm.
The resulting expression for the collision rate $1/\tau$ is evaluated numerically. We note that it depends only  
on $T/T_F$  because rescaling the $x,y$ coordinates by the harmonic oscillator 
lengths $l_{x,y}=1/\sqrt{m \omega_{x,y}}$ cancels any additional dependence on trap frequencies. 
The fugacity $z$ is found from to $T/T_F$ via the formula for
the non-interacting Fermi gas in a harmonic trap $T_F/T=\sqrt{-2 {\rm Li}_2(-z)}$.

In Fig.~\ref{fig:fig1} (b), we plot the collision rate as a function of interaction 
strength $\ln(k_Fa_{\rm 2D})$ for $T/T_F=0.47$ and as a function of $T/T_F$ for $\ln(k_{\rm F} a_{\rm 2D})=0.5$
using three different approximations: The blue dashed curves are the classical expression given by Eq.~(\ref{tauclass}),
the black solid curves include Pauli blocking in the collision integral, and the red dotted curves include Pauli 
blocking as well as the medium corrections to the ${\mathcal T}$-matrix given by Eq.~(\ref{eq:mbtmatrix}).
The classical collision rate monotonically increases as the temperature is lowered as a consequence 
of the increasing density. 
 When Pauli blocking is included this behavior is drastically different, since the exclusion principle freezes out collisions in 
 the degenerate regime $T \ll T_F$. This freezing out results in a maximum of the collision rate as a function 
 of temperature  as shown in the inset of Fig. \ref{fig:fig1} (b).

This picture is further modified when medium corrections in the ${\mathcal T}$-matrix are included. 
Here the proximity to the BCS instability towards pair formation gives rise to resonant behavior for particles colliding 
with opposite momenta near the Fermi surface. In the strong coupling regime $|\ln(k_{\rm F} a_{\rm 2D})|\lesssim 2$, 
 this causes a dramatic 
increase in the scattering rate over a large temperature range, diverging at the mean-field BCS 
transition temperature $T^*$ given by the Thouless criterion, as indicated by the dashed vertical lines in Fig. \ref{fig:fig1} (b).
While there is strictly speaking no phase transition at $T^*$ in 2D, this should be considered as the temperature where pairing sets in, but fluctuations prevent superfluid order as long as $T$ is above the Berezinskii-Kosterlitz-Thouless temperature $T_{\rm BKT}$~\cite{Loktev:2001fk,Feld:2011nx}.
The medium corrections  in fact  over-compensate the Pauli blocking effects for strong coupling resulting in 
a collision rate which is \emph{larger} than the classical prediction, as can be seen from Fig. \ref{fig:fig1} (b). 
This over-compensation  is consistent with what was found for 
the shear viscosity and spin diffusion coefficient~\cite{Enss:2012ly}. 
Medium corrections to the scattering in the strong coupling
 regime  are thus even larger in 2D than in 3D~\cite{Riedl:2008bh}, which can be interpreted as showing the presence of a 
large so-called pairing pseudo-gap regime above the superfluid transition temperature $T_{\rm BKT}$~\cite{Loktev:2001fk}. 

On the molecular side, sufficiently far beyond the BCS transition, 
the collision rate is again finite. This is because  the pole of the ${\mathcal T}$-matrix is below the integration region in the collision integral, where 
the ${\mathcal T}$-matrix is evaluated on shell for the collision process. 
 In this limit, our theory effectively describes a repulsive Fermi gas in the upper branch~\cite{Shenoy:2011ys}.     
Away from the BCS transition, the collision rate with medium effects approaches the Pauli-blocked result.

\section{Comparison with experiment}
We now compare our theoretical results  with the experimental data of Ref.~\cite{Vogt:2012vn}. 
We focus on the experimental values for the temperature $T/T_F=0.47$ where the available data had the best signal-to-noise ratio. 
Also, analysis techniques have been adjusted and refined for this temperature, which marginally changed the experimental data points 
without changing the overall results and statements of Ref.~\cite{Vogt:2012vn}.
It should be emphasized that the theoretical expressions for the frequency and damping contain no
free parameters to fit theory and experiment. In Figs.~\ref{fig:fig2}  (a) and ~\ref{fig:fig2} (b) we plot 
the frequency and damping of the quadrupole mode. As for the collision rate, we show three different 
curves: A classical  calculation (blue dashed curve), a calculation including Pauli blocking (black solid curve), and a calculation 
where both Fermi blocking and medium effects on the ${\mathcal T}$-matrix are included (red dotted curve). We see that 
there is good  agreement with the experimental results for 
quadrupole frequency and damping rates. In particular, the onset of the hydrodynamic 
 regime around $\ln (k_{\rm F} a_{\rm 2D}) \sim 2-5$ and the location of the maximum damping rate seen in the experiment agree with our 
 calculations. 

We see from Fig.~\ref{fig:fig1} that the collision rate including medium corrections in the scattering matrix is in between the classical result  and the 
Pauli blocking result for $\ln(k_Fa_{\rm 2D})\gtrsim1.6$ and $T/T_F=0.47$.
 This explains why the frequency in Fig.~\ref{fig:fig2} is in between the classical and the Pauli blocking 
predictions in this regime. When $-0.6\lesssim\ln(k_Fa_{\rm 2D})\lesssim1.6$ the medium collision rate is larger than the classical prediction, but in this regime the mode frequency  is already 
very close to the hydrodynamic value, so there is no visible difference between the classical and the medium prediction for the frequency. The higher collision rate when medium corrections 
are included does, however, yield a smaller damping in the hydrodynamic regime, as can be see from Fig.~\ref{fig:fig2} (b).  
  Somewhat surprisingly, it appears  that including medium corrections gives a slightly worse agreement with the 
 experimental results compared to the theory that includes only Pauli blocking.  
Experimentally,
when tuning from attractive interactions beyond $\ln(k_{\rm F} a_{\rm 2D})\lesssim 1$ one observes rapid heating of the gas, qualitatively in agreement with the expectation of increased pairing correlations yielding a higher three-body loss rate when entering the pairing pseudo-gap region $T<T^*$.
We note that our theoretical approach is not valid in the
 paired regime, and further theoretical studies are needed to fully address the collective modes at low temperatures.

\begin{figure*}
\includegraphics[width=2.09\columnwidth]{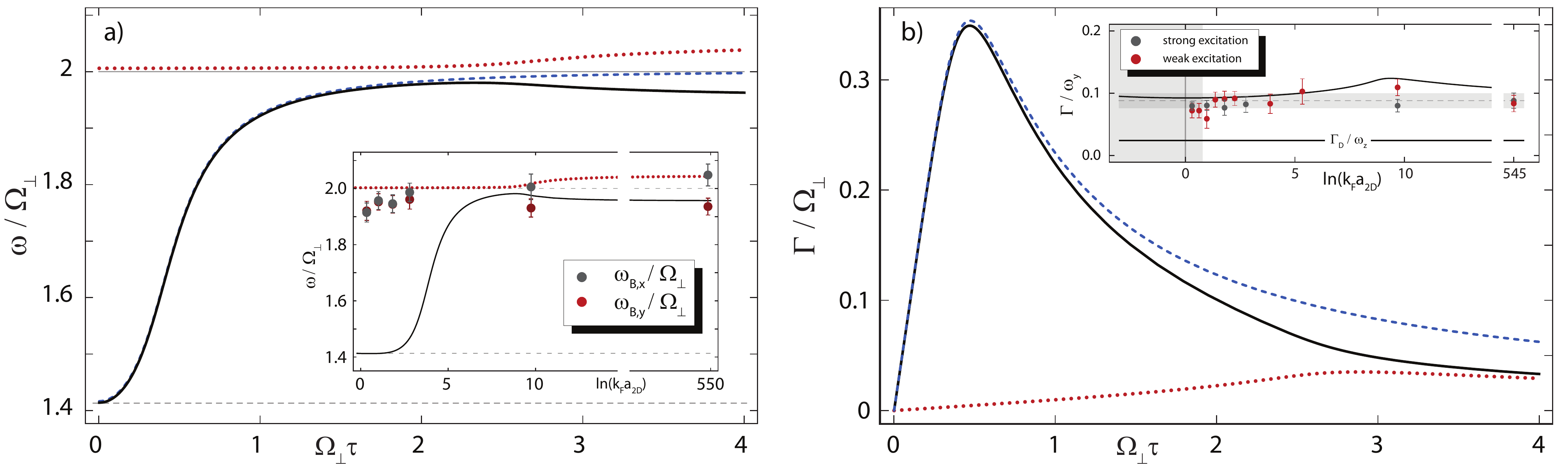}
\caption{(Color online) Frequency (a) and damping (b) of the lowest collective modes of a two component Fermi gas in a slightly anisotropic quasi-2D harmonic trap ($\omega_y=1.05 \omega_x$) normalized by the geometric mean of the frequencies $\Omega_{\perp}=(\omega_x \omega_y)^{1/2}$ 
 as a function of collision time $\tau$.  The thin gray lines in (a) denote the isotropic limits of the breathing and quadrupole mode frequencies $\omega_{B, Q}$ in the collisionless $\omega_{B,Q}/\Omega_{\perp} \equiv 2$ and the hydrodynamic regimes with $\omega_{B}/\Omega_{\perp} \equiv 2, \omega_{Q}/\Omega_{\perp} = \sqrt{2}$, respectively.
In an anisotropic trap, the breathing (dotted red line) and quadrupole (solid black line) modes are coupled. The blue dashed curves correspond to the scissors mode (see the Appendix). 
Insets: (a) Decoupled eigenmodes in the collisionless regime [$\ln(k_Fa_{\rm 2D})\gg10$]
start to lock towards the strongly-interacting regime [$\ln(k_Fa_{\rm 2D})\lesssim5$], resulting in a single collective mode with strong breathing mode character.
(b) Same plots for the parameters of the experiment ~\cite{Vogt:2012vn} as a function of interaction strength at $T=0.47\,T_F$ (here shown for the collision time obtained from the theory with Pauli blocking). The dots are the experimental data for the breathing-mode damping rate for two different excitation strengths~\cite{Footnote1}.} 
\label{fig:normalmodes}
\end{figure*}

\section{Anisotropic trap}
It is typically quite difficult to achieve perfect symmetric traps in experiments. In the experiment of Ref. ~\cite{Vogt:2012vn}, there is some  
 uncertainty in the value of the trapping frequencies $\omega_x$ and $\omega_y$ so that the trap  may not be perfectly isotropic in the $xy$-plane. 
 Instead, one typically has an upper bound 
of $|\omega_x/\omega_y-1|\lesssim 5\%$  for the anisotropy. 
An anisotropic trap results in coupled breathing  and quadrupole modes~\cite{Guery-Odelin:1999kx,Ghosh:2002vn}, and we shall now show that 
this can have important consequences on the interpretation of the experimental results. 

We consider an anisotropic trap with $\omega_x \neq \omega_y$. To describe the coupled breathing and quadrupole modes, we 
use a generalized version of the basis functions Eqs. (\ref{eq:basis1}-\ref{eq:basis2}):
\begin{eqnarray}
\label{eq:qp1}
\phi_1&=&\omega_x^2 x^2+\omega_y^2 y^2, \; \phi_2= \omega_x x p_x+\omega_y y p_y, \\ \label{eq:qp2} \phi_3&=&p_x^2+p_y^2, \; \phi_4= \omega_x^2 x^2-\omega_y^2 
y^2,\\ \label{eq:qp3}
\phi_5&=&\omega_x x p_x-\omega_y y p_y,\; \phi_6=p_x^2-p_y^2.
\end{eqnarray}
The calculation of the collective modes using this basis set is described in the Appendix. After a straightforward but rather lengthy calculation, 
we obtain that the normal mode frequencies are the solutions of
\begin{equation}
\label{eq:normalmodes}
(\omega^2-\omega_{\rm cl1}^2) (\omega^2-\omega_{\rm cl2}^2)+\frac{i}{\omega \tau} (\omega^2-\omega_{\rm hd1}^2) (\omega^2-\omega_{\rm hd2}^2)=0
\end{equation}
with $\omega_{\rm cl1/2}=2 \omega_{x/y}$ the collisionless frequencies. The hydrodynamic frequencies are 
\begin{equation}
\omega_{\rm hd,1/2}=\sqrt{\frac{3}{2}} \sqrt{\omega_x^2+\omega_y^2\pm \sqrt{\omega_x^4- \frac{14}{9} \omega_x^2 \omega_y^2+\omega_y^4}}.
\end{equation} 
The collision time $\tau$ appearing in Eq. \eqref{eq:normalmodes} is identical to the collision time for an isotropic trap, 
when evaluated at the same fugacity $z=e^{\beta \mu}$, since we can eliminate any dependence on trap frequencies by rescaling the real-space coordinates $x$,$y$.

In Fig. \ref{fig:normalmodes} we show  the frequencies and damping rates of these modes as a function of the collision time $\tau$ for 
$\omega_x=1.05\,\omega_y$. The coupling of the breathing and quadrupole modes is clearly visible. In the insets of Fig. \ref{fig:normalmodes}, we compare our theory with the experimental data for the frequency and damping of the breathing mode in Ref.~\cite{Vogt:2012vn}. The inset of Fig. 3 (a) shows the separately measured cloud-width oscillation frequencies in the $x$, and $y$ directions of the breathing mode as a function of interaction strength. In the collisionless regime ($\ln(k_Fa_{\rm 2D})\gg1$) the oscillations along the principal axes of the system are decoupled. The frequency difference directly reveals the system's anisotropy of about 5\%. Towards the hydrodynamic regime the motion along the main axes 
start to couple, leading to a single collective mode in the strong coupling limit [$\ln(k_Fa_{\rm 2D})\lesssim3$]. The breathing mode frequency in that regime lies below the theoretical prediction, probably due to the anisotropy, which couples the monopole mode with the quadrupole mode.
An  important consequence of this coupling is that there is a viscous damping of the breathing mode, which is  undamped for an isotropic harmonic trap as the kinetic theory yields zero
bulk viscosity. This potentially complicates the interpretation of the observed damping of the breathing mode in the experiment of Ref. \cite{Vogt:2012vn}, since it is 
necessary to separate the damping due to non-zero shear viscosity caused by a possible trap anisotropy from damping due to non-zero bulk viscosity. 
In fact, the  inset of Fig. \ref{fig:normalmodes} (b) demonstrates that a small trap anisotropy of $5$-$10$\%  can qualitatively explain 
the slight increase in damping of the breathing mode~\cite{Footnote1} with coupling strength observed in Ref. \cite{Vogt:2012vn}.  
This additional source of damping makes the bulk viscosity extracted from the damping even smaller,
and more measurements are needed in order to  
pin down how small the  bulk viscosity is and to compare with theory~\cite{Hofmann:2012fk,Taylor:2012kx,Gao:2012uq}.

\section{Conclusions}
We studied the collective modes of a 2D Fermi gas using kinetic theory. Expanding the Boltzmann equation on basis functions, we 
obtained analytical results for the frequency and damping of the collective modes as a function of temperature and coupling strength, which were 
shown to agree  well with experimental data. We demonstrated that there is a large temperature range for strong coupling, where pairing correlations 
dominate the effects of Pauli blocking on the collision rate, resulting in a rate higher than the classical value. Finally, we showed that the coupling 
between the breathing and the quadrupole modes due to a slight anisotropy within experimental bounds, may result in a damping of the 
breathing mode even when the bulk viscosity is zero. This can have important consequences for the interpretation of the experimental results
for the breathing mode.

\acknowledgements
This work has been supported by the EPSRC Grants No. EP/I010580/1 (S.B.),  No. EP/J0149X/1 and No. EP/K003615/1 (M. K. and E. V.). M. K. and E. V. acknowledge support from the Royal Society and the Wolfson Foundation. G. M. B. acknowledges support from the Carlsberg Foundation.
\newpage
\onecolumngrid
\appendix
\section{Derivation of the equations for the collective modes in an anisotropic trap}
\label{sec:app1}
To find the collective modes, we expand the linearized Boltzmann equation~\eqref{eq:linearizedboltzmann} in a finite set of basis functions. This allows us to obtain simple matrix equations for the collective mode frequencies and damping rates. Formally we are interested in solutions of the equation $\hat{{\cal L}}_{\omega} \phi=0$ with the linear operator $\hat{{\cal L}}_{\omega}$ defined as
\begin{equation}
\label{eq:linearboltzmann2}
\hat{{\cal L}}_{\omega} \phi=f_0 (1-f_0) \left(-i \omega+ \dot{\mathbf{r}} \cdot \mathbf{\nabla_{\mathbf{r}}}+\dot{\mathbf{p}} \cdot \mathbf{\nabla_{\mathbf{p}}} \right) \phi+I[\phi].
\end{equation}
To solve this equation we project Eq.~\eqref{eq:linearboltzmann2} onto a set of functions $\phi(\mathbf{r},\mathbf{p})=\sum_i c_i \phi_i(\mathbf{r},\mathbf{p})$, converting the linearized Boltzmann equation into a tractable finite dimensional matrix equation of the form
\begin{eqnarray}
\sum_j M_{ij}(\omega) c_j=0
\end{eqnarray}
with coefficients given by
\begin{eqnarray}
M_{ij}(\omega)=\frac{\int d^2r d^2p\; \phi_i \hat{{\cal L}}_{\omega} \phi_j}{\int d^2r d^2p f_0 (1-f_0) \phi_i^2}.
\end{eqnarray}
The collision integral $I[\phi]$ vanishes for any function
\begin{equation}
\label{eq:vanishingcintegral}
\phi(\mathbf{r},\mathbf{p})=a(\mathbf{r})+\mathbf{p} \cdot \mathbf{b}(\mathbf{r})+\mathbf{p}^2 c(\mathbf{r})
\end{equation}
since it is local in real space and because both momentum and energy are conserved during collisions~\cite{PethickSmith}.
Applying this formalism to the moments Eqs. (\ref{eq:qp1})- (\ref{eq:qp3}) yields then the set of equations for a general harmonic trap
\begin{equation}
\label{eq:matrixequation}
\left(
\begin{array}{cccccc}
 -i \omega  & \frac{1}{4} \left(\omega _x+\omega _y\right) & -\frac{i \omega }{2} & 0 & \frac{1}{4} \left(\omega _x-\omega _y\right) &
   0 \\
 -(\omega _x+\omega _y) & -i \omega  & \omega _x+\omega _y & \omega _y-\omega _x & 0 & \omega _x-\omega _y \\
 -\frac{i \omega }{2} & -\frac{1}{4} \left(\omega _x+\omega _y\right) & -i \omega  & 0 & \frac{1}{4} \left(\omega _y-\omega _x\right)
   & 0 \\
 0 & \frac{1}{2} \left(\omega _x-\omega _y\right) & 0 & -i \omega  & \frac{1}{2} \left(\omega _x+\omega _y\right) & 0 \\
 \omega _y-\omega _x & 0 & \omega _x-\omega _y & -(\omega _x+\omega _y) & -i \omega  & \omega _x+\omega _y \\
 0 & \frac{1}{2} \left(\omega _y-\omega _x\right) & 0 & 0 & -\frac{1}{2} \left(\omega _x+\omega _y\right) &-i \omega + \frac{1}{\tau }
   \\
\end{array}
\right)
\left(
\begin{array}{c}
c_1 \\
c_2 \\
c_3 \\
c_4 \\
c_6
\end{array}
\right)
=0
\end{equation}
This matrix equation has nontrivial solutions when the coefficient determinant vanishes. This results in Eq.~\eqref{eq:normalmodes} for the normal-mode frequencies of the coupled monopole and quadrupole modes described by the basis functions Eqs. (\ref{eq:qp1})-(\ref{eq:qp3}). Here the only basis function for which the collision integral does not vanish is $\phi_6=p_x^2-p_y^2$. By parity symmetry only the matrix element $M_{66}$ receives a contribution $1/\tau$ given by Eq.~\eqref{eq:relaxationtime}, giving a finite damping rate to the collective-mode frequencies.  For an isotropic trap, Eq. ~\eqref{eq:matrixequation} decouples into an independent undamped breathing at twice the trapping frequency~\cite{Guery-Odelin:1999kx} and a damped quadrupole mode with frequency and damping given by Eq.~\eqref{eq:qpmode}. Finally, we mention that independent of the coupled quadrupole and breathing modes, there is also the  scissors mode 
which can be described by the basis functions~\cite{Bruun:2007ys}
\begin{eqnarray}
\label{eq:scissorsbasis}
\phi_7&=&\omega_x \omega_y x y, \phi_8=\omega_x p_y x+\omega_y p_x y,\\ \phi_9&=&\omega_x p_y x-\omega_y p_x y, \phi_{10}=p_x p_y
\end{eqnarray}
In the isotropic case $\omega_x=\omega_y$, this mode is simply one of the degenerate rotations of the quadrupole mode. The frequencies of the scissors mode are the solutions of the equation~\cite{Guery-Odelin:1999ys,Bruun:2007ys}
\begin{equation}
\label{eq;scissors}
\frac{i \omega}{\tau'}(\omega^2-\omega_{\rm hd}'^2)+(\omega^2-{\omega_{\rm cl1}'}^2)(\omega^2-{\omega_{\rm cl2}'}^2)=0
\end{equation}
where $\omega_{\rm hd}'=\sqrt{\omega_x^2+\omega_y^2}$ is the frequency of the scissors mode in the hydrodynamic limit, whereas 
 $\omega_{\rm cl1}'=\omega_x+\omega_y$ and  $\omega_{\rm cl2}'=|\omega_x-\omega_y|$ are the mode frequencies in the collisionless limit. One obtains the matrix equation for the scissors mode
\begin{equation}
\left(
\begin{array}{cccc}
 -i \omega  & \omega _x+\omega _y & \omega _y-\omega _x & 0 \\
 -\frac{1}{2} \left(\omega _x+\omega _y\right) & -i \omega  & 0 & \frac{1}{2} \left(\omega _x+\omega _y\right) \\
 \frac{1}{2} \left(\omega _x-\omega _y\right) & 0 & -i \omega  & \frac{1}{2} \left(\omega _x-\omega _y\right) \\
 0 & -(\omega _x+\omega _y) & \omega _y-\omega _x &-i \omega+ \frac{1}{\tau'}  \\
\end{array}
\right)
\left(
\begin{array}{c}
c_7 \\
c_8 \\
c_9 \\
c_{10}
\end{array}
\right)
=0
\end{equation}
when projecting onto the basis functions Eq.~\eqref{eq:scissorsbasis}. Again this equation has nontrivial solutions when the determinant vanishes, leading to Eq.~\eqref{eq;scissors}. In the isotropic limit, only three basis functions are coupled and in this case they simply describe the quadrupole mode [rotated by 45$^\circ$ with respect to the quadrupole mode from Eq.~\eqref{eq:matrixequation}]. The collision rate for the scissors mode is given by 
\begin{eqnarray}
\frac{1}{\tau'}=\frac{\int d^2r d^2\check{p}\; p_x p_y I[p_x p_y]}{\int d^2r d^2\check{p} f_0(1-f_0) p_x^2 p_y^2}.
\end{eqnarray}
Since both $p_xp_y$ and $p_x^2-p_y^2$ belong to the $l=2$ representation of the rotation group, we have $\tau'=\tau$. This result is also valid in the more general anisotropic case where $\omega_x\neq \omega_y$.
\twocolumngrid

\end{document}